\documentclass[conference]{IEEEtran}
\IEEEoverridecommandlockouts

\usepackage{cite}
\usepackage{amsmath,amssymb,amsfonts}
\usepackage{algorithmic}
\usepackage{graphicx}
\usepackage{textcomp}
\usepackage[dvipsnames]{xcolor}

\usepackage{booktabs} 
\usepackage[ruled,linesnumbered]{algorithm2e}
\def\BibTeX{{\rm B\kern-.05em{\sc i\kern-.025em b}\kern-.08em
    T\kern-.1667em\lower.7ex\hbox{E}\kern-.125emX}}

\usepackage{pifont}

\usepackage{array}    
\newcolumntype{L}[1]{>{\raggedright\let\newline\\\arraybackslash\hspace{0pt}}m{#1}}
\newcolumntype{C}[1]{>{\centering\let\newline\\\arraybackslash\hspace{0pt}}m{#1}}
\newcolumntype{R}[1]{>{\raggedleft\let\newline\\\arraybackslash\hspace{0pt}}m{#1}}

\IEEEpubid{\makebox[\columnwidth]{978-1-6654-3274-0/21/\$31.00 ©2021 IEEE \hfill} \hspace{\columnsep}\makebox[\columnwidth]{ }}

\begin{document}

\title{ GNN4IP: Graph Neural Network for Hardware Intellectual Property Piracy Detection 
\thanks{* Both are equal contributing authors. Yasaei is the corresponding author.}
}
\author{\IEEEauthorblockN{Rozhin Yasaei*, Shih-Yuan Yu*, Emad Kasaeyan Naeini, Mohammad Abdullah Al Faruque}
\IEEEauthorblockA{\textit{Department of Electrical Engineering and Computer Science} \\
\textit{University of California, Irvine, California, USA} }}

\maketitle

\begin{abstract}

Aggressive time-to-market constraints and enormous hardware design and fabrication costs have pushed the semiconductor industry toward hardware Intellectual Properties (IP) core design. However, the globalization of the integrated circuits (IC) supply chain exposes IP providers to theft and illegal redistribution of IPs.  Watermarking and fingerprinting are proposed to detect IP piracy. Nevertheless, they come with additional hardware overhead and cannot guarantee IP security as advanced attacks are reported to remove the watermark, forge, or bypass it. In this work, we propose a novel methodology, GNN4IP, to assess similarities between circuits and detect IP piracy. We model the hardware design as a graph and construct a graph neural network model to learn its behavior using the comprehensive dataset of register transfer level codes and gate-level netlists that we have gathered. GNN4IP detects IP piracy with 96\% accuracy in our dataset and recognizes the original IP in its obfuscated version with 100\% accuracy.

\end{abstract}


\vspace{-1em}
\section{Introduction}

The integrated circuits (IC) manufacturing industry has developed significantly and scaled down to 7nm technology that has made the integration of numerous transistors possible. However, the hardware engineers could not keep up with rapid advancement in the fabrication technology, and they fail to use all of the available transistors in the die. To close this productivity gap under time-to-market pressure, hardware Intellectual Property (IP) core design has grabbed substantial attention from the semiconductor industry and has dramatically reduced the design and verification cost \cite{chen2020decoy}.

The globalization of the IC supply chain poses a high risk of theft for design companies that share their most valuable assets, IPs, with other entities. IP piracy is a serious issue in the current economy, with a drastic need for an effective detection method. According to the U.S. Department of Commerce study, 38\% of the American economy is composed of IP-intensive industries \cite{news1} that lose between \$225 billion to \$600 billion annually because of Chinese companies stealing American IPs mainly in the semiconductor industry, based on the U.S. Trade Representative report \cite{news2}.

Hardware IP is considered as any stand-alone component of a system-on-chip design that is classified into three categories based on the level of abstraction: Soft IP (i.e., synthesizable HDL source code), Firm IP (i.e., netlists and placed RTL block), and Hard IP (i.e., GDSII and physical layout) \cite{chang2016hardware}. 

Conventionally, the IP protection techniques fall into preventive (i.e., logic encryption, camouflaging, metering, and split manufacturing) and detective (i.e., digital signature) methods. All these methods add excessive implementation overhead to the hardware design that limits their applications in practice. Moreover, they mainly focus on security at the IC level, while many commercial IPs comprise the soft IPs due to flexibility, independence of platform technology, portability, and easy integration with other components. The high level of abstraction makes the IP protection more challenging since it is easier for an adversary to slightly change the source code and redistribute it illegally at the lower levels of abstraction. Although the existing preventive countermeasures deter IP theft, they cannot guarantee IP security as the adversaries keep developing more sophisticated attacks to bypass them. \textbf{Therefore, an effective IP piracy detection method is crucial for IP providers to disclose the theft.} 

To this end, the state-of-the-art piracy detection method is embedding signatures of IP owner known as watermark and legal IP user known as a fingerprint in the circuit design to assure authorship and trace legal/illegal IP usage. IP watermarking and fingerprinting are prone to removal, masking, or forging attacks that attempt to omit the watermark, distort its extraction process, or embed another watermark in IP \cite{chang2016hardware}. 

\textbf{In this work, we propose a novel methodology for IP piracy detection that, instead of insertion and extraction of a signature to prove the ownership, models the circuits and assess the similarity between IP designs. Therefore, our method does not require additional hardware overhead as the signature and is not vulnerable to removal, masking, or forging attacks. It also effectively expose the infringement between two IPs when the adversary complicates the original IP to deceive the IP owner.
Modeling the hardware design is challenging since it is a structural non-Euclidean data type, despite most modeling techniques. Thus, similar to \cite{gnn4tj}, we represent the circuit as a data-flow graph (DFG)  due to similar data types and properties. Afterward, we model it using state-of-the-art graph learning method.}

\vspace{-0.5em}
\subsection{Motivational Example}

\begin{figure}[t]
    \centering
    \includegraphics[width=0.48\textwidth]{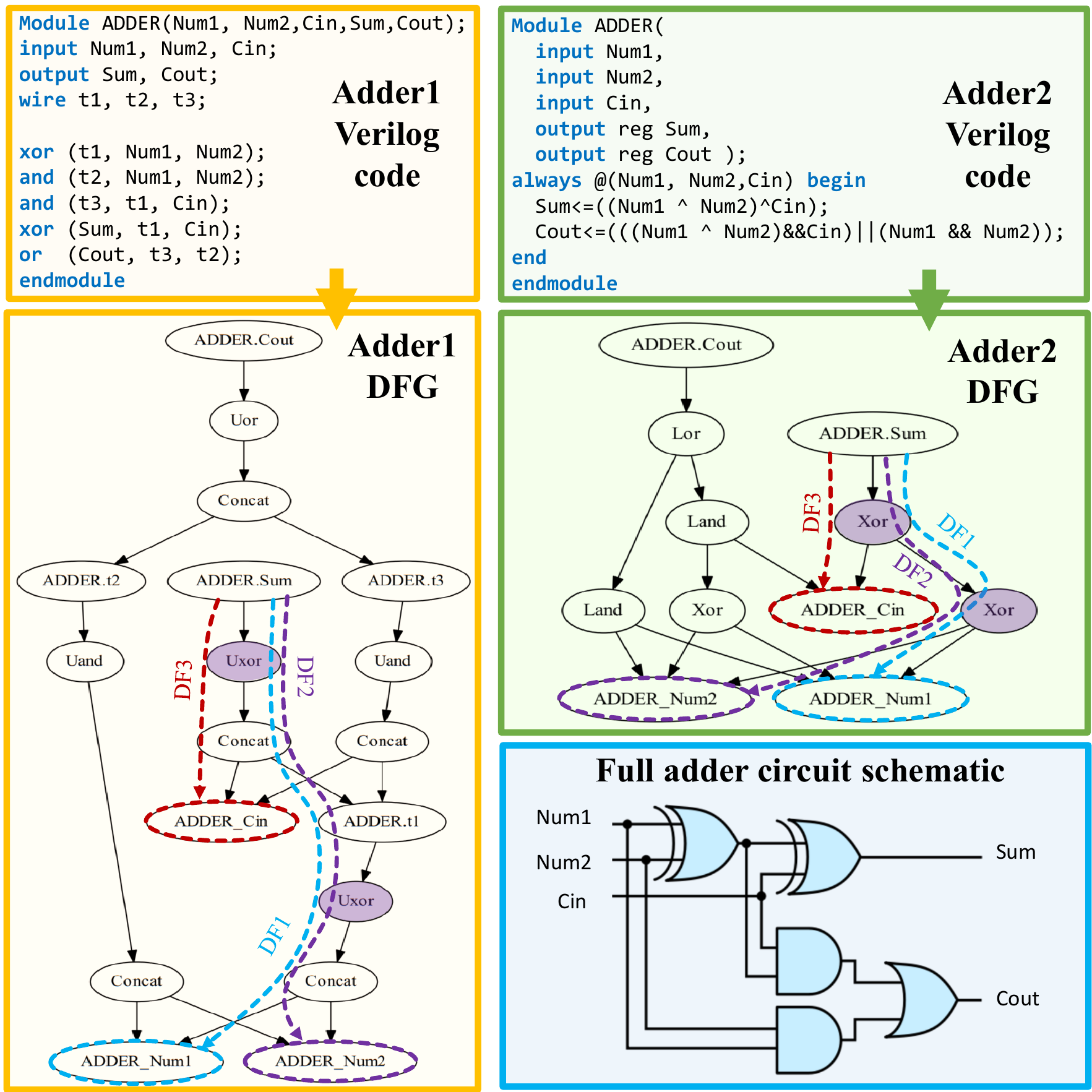}
    \vspace{-1em}
    \caption{The circuit schematic, Verilog codes, and DFGs of full adder circuits.}
    \label{fig:example}
    \end{figure}

We study the concept of piracy and similarity among hardware designs in a test case of two different variations of the full adder circuit. As shown in Figure \ref{fig:example}, although the Verilog codes for adder 1 and 2 are different, they both have fundamentally the same design, as depicted in the schematic figure. We unveil the similarity between two adders using DFG, which expresses the signals dependency and computational structure. At first glance, the generated DFGs for adders seem varied, but a deep look into data flows (DF) indicates the same signal relations. For instance, the output signal \textit{Sum} depends on \textit{Num1}, \textit{Num2}, and \textit{Cin} input signals through the DF1, DF2, and DF3, respectively. Suppose we focus on critical nodes in the flow (XOR nodes) and ignore the excessive nodes related to concatenation and internal signals. In that case, the DFs in both DFGs represent the same operations. 

\vspace{-0.8em}
\subsection{Research Challenges}

The development of an effective IP piracy detection method poses paramount research challenges as follows:
 \begin{itemize}
    \item \textbf{Hardware overhead}: All existing piracy detection methods add hardware overhead to IP design.
    \item \textbf{Attacks}: Signatures-based countermeasures are vulnerable to removal, forging, and masking attacks.
	\item \textbf{Same behaviors, different topologies}: As the case study exemplified, varied HDL codes generate different DFGs even if they represent the same hardware design. The different typologies in DFGs can easily fool the standard graph similarity algorithms, and behavioral analysis of graphs is required to learn circuit design. 
	\item \textbf{Scalability}: The manual review of hardware design is not feasible in practice. The graph similarity is an NP-complete problem, and existing algorithms \cite{fyrbiak2019graph} suffer from high complexity and are not scalable to large designs and industrial-level IPs with thousands of code lines.
\end{itemize}


\vspace{-0.8em}
\subsection{Our contribution}

We propose a novel methodology based on graph learning to surmount research challenges and propose these contributions:

\begin{itemize} 
	 \item To overcome the shortcomings of current IP piracy detection methods, we propose a novel countermeasure based on hardware design analysis that does not require adding any signature and overhead to IP design.
	\item We develop a scalable, automated framework called \textit{hw2vec} that generates the DFG for hardware designs and assigns a embedding to them such that the proximity in the embeddings indicates similarity between circuits.
	\item We construct a Graph Neural Network (GNN) model to learn the circuit's behavior and assess the similarity between a pair of IPs according to graph embeddings.
	\item We gather a dataset of hardware designs in RTL and gate-level netlist to develop and assess our methodology.
\end{itemize}

\vspace{-0.8em}
\section{Backgrounds and Related Works}

\vspace{-0.2em}
\subsection{Hardware IP Security}


The hardware is susceptible to security threats such as IP piracy (unlicensed usage of IP), overbuilding, counterfeiting (producing a faithful copy of circuit), reverse engineering, hardware Trojan (malicious modification of circuit) \cite{htm, sina1, gnn4tj}, and side-channel attacks \cite{ashrafiamiri2018towards}. 
The IP protection methods proposed in the literature can be classified as follows:

\textbf{Watermarking and fingerprinting} \cite{rai2019hardware}: The IP owner and legal IP user's signatures, known as watermark and fingerprint, are added to circuit to prove infringement. 
    
\textbf{Hardware metering} \cite{koushanfar2017active}: The designer assigns a unique tag to each chip, which can be used for chip identification (passive tag) or enabling/disabling the chip (active tag). 
    
\textbf{Obfuscation} \cite{chen2020decoy}: There are two obfuscation methodologies; \textbf{logic locking} (encryption) \cite{xie2017delay} and \textbf{IC camouflaging} \cite{camouflaging}. In logic locking, additional gates such as XOR are inserted in non-critical wires. The circuit would be functional only if the correct key is provided which is stored in a secure memory out of reach of the attacker. Camouflaging modifies the design such that cells with different functionalities look similar to the attacker and confuses the reverse engineering process. 

\textbf{Split manufacturing} \cite{patnaik2018raise}: IP house split the design to separate ICs and have them fabricated in different foundries. Thus, none of the foundries have access to the whole design to overbuild, reverse engineer, or perform  malicious activities. 

The existing defenses suffer from a large overhead on area, power, and timing that restrict their application. As the new countermeasures are developed, the attacks are advanced to bypass them. SAT attack is a powerful method used to formulate and solve a sequence of SAT formulas iteratively to unlock the encrypted circuit, reverse engineer the Boolean functionalities of camouflaged gates \cite{el2019sat}, or reconstruct the missing wire in 2.5D split manufactured ICs \cite{wang2019reverse}. 
Anti-SAT \cite{xie2016mitigating}, and AND-tree insertion \cite{li2017provably} obfuscation techniques are proposed to mitigate SAT attack. However, signal probability skew attack, AppSAT guided removal attack, and sensitization guided SAT attack \cite{yasin2017removal} break them. Proximity attack \cite{rajendran2013split} is another attack against 2.5D split manufacturing that iteratively connects the inputs to outputs in two IC partitions until a loop is formed. Removal, masking, and forging attacks bypass watermarking by eliminating, distorting, or embedding a ghost watermark \cite{chang2016hardware}. There is a rising trend in machine learning-based defenses \cite{gnn4tj, yasaei2020iot} and the recent advances made the models even resistant against adversarial attacks \cite{ashrafiamiri2020r2ad}.

\vspace{-0.5em}
\subsection{Graph Neural Networks (GNNs)} 
\label{subsec:gnn}

In \textit{GNN4IP}, we leverage GNN, a deep learning methodology that tackles graph data~\cite{wu2020comprehensive}. Several works in the literature have shown the effectiveness of GNN in identifying software clones and detecting binary code similarity~\cite{9054857,xu2017neural}.
Our architecture is inspired by the Spatial-based Graph Convolution Neural Network, which defines the convolution operation based on a node's spatial relations with the following phases: (i) \textit{message propagation phase} and (ii) the \textit{read-out phase}. 
The \textit{message propagation} phase involves two sub-functions: \textbf{AGGREGATE} and \textbf{COMBINE}, given by,
\begin{equation}
    \vspace{-0.5em}
    a_v^{(k)} = \textbf{AGGREGATE}^{(k)}(\{h_u^{(k-1)}: u \in N(v)\}),
\end{equation}
\begin{equation}
    h_v^{(k)} = \textbf{COMBINE}^{(k)}(h_v^{(k-1)}, a_v^{(k)} ),
    \vspace{-0.5em}
\end{equation}
where $h_v^{(k)} \in R^{C^k}$ denotes the node embedding after $k$ iterations for the $v_{th}$ node. 
Essentially, the \textbf{AGGREGATE} function collects the features of the neighboring nodes to extract an aggregated embedding $a_v^{(k)}$ for the layer $k$, and the \textbf{COMBINE} function combines the previous node features $h_v^{(k-1)}$ with  $a_v^{(k)}$ to output next embedding $h_v^{(k)}$. 
This message propagation is carried out for a pre-determined number of iterations $k$.
Next, in the \textit{read-out} phase, the overall graph-level embedding extraction is carried out by either summing up or averaging up the node embeddings in each iteration. 
The graph-level embedding is denoted as $h^{(k)}_G$ and is defined as,
\begin{equation}
    \vspace{-0.5em}
    h^{(k)}_G = \textit{READOUT}(\{h_v^{(k-1)}: v \in G\})
\end{equation}
In our work, we use $h^{(k)}_G$ as the hardware design embedding to assess the similarity between circuits and discover piracy.

\vspace{-0.5em}
\section{GNN4IP Methodology}

In this work, we formulate the problem of IP piracy detection as finding the similarity between two hardware designs.
We assume the existence of a feed-forward function $f$ that outputs whether two circuits $p_A$ and $p_B$ are subject to piracy or not through a binary label $y$ as given in Equation~\ref{formular1}.
\begin{equation}
\label{formular1}
    y = f(p_A, p_B) = \left \{
    \begin{array}{ll}
         (1,0) & \text{if piracy in $p_A, p_B$}\\
         (0,1) &  \text{if no-piracy in $p_A, p_B$}
    \end{array}
    \right.
\end{equation} 

To approximate $f$, we extract the DFGs $G_A$ and $G_B$ from circuits pair $(p_A, p_B)$ using the \textit{DFG generation pipeline} and pass it to a graph embedding layer, \textit{hw2vec}, to acquire embeddings $(h_{G_A}, h_{G_B})$. Lastly, our model infer the piracy label, $\hat{Y}$, by computing the cosine similarity between $(h_{G_A}, h_{G_B})$.

\vspace{-0.5em}
\subsection{Threat Model}

In our threat model, we examine the IP designs in RTL or gate-level netlist to discover piracy. We assume that the design is a soft IP, firm IP, or derived by reverse engineering a hard IP or IC. 
The adversary can be a hardware designer, competitor company, or the fabrication foundry who present the stolen IP as their genuine design and sell it as an IC or an IP at the same or lower level of abstraction. The attack scenario may involve modification of IP design to tamper piracy detection. The attacker can get access to the original IP through one of these means: I) purchase the IP for limited usage, II) leaked through a rogue employee in the design house, or III) reverse engineered the physical layout or IC. 

    \begin{figure}[t]
    \centering
    \includegraphics[width=0.4\textwidth]{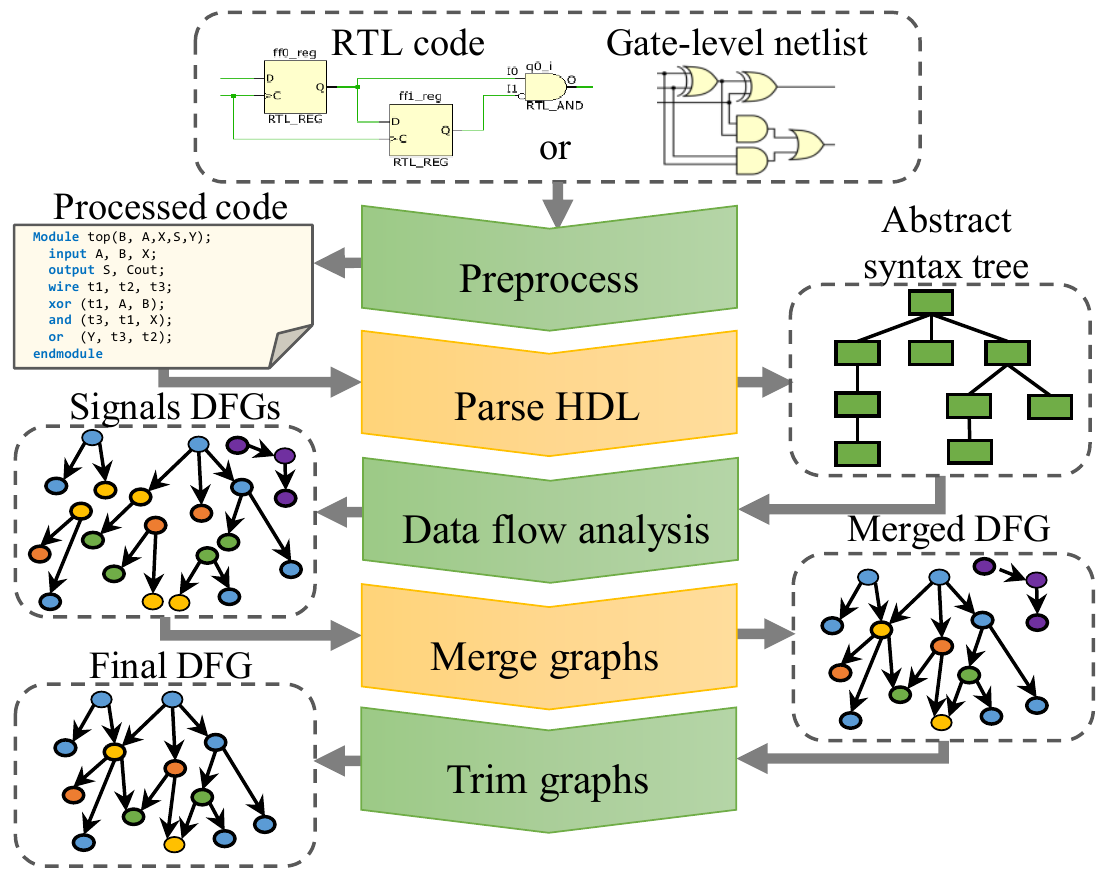}
    \vspace{-1em}
    \caption{Data flow graph generation pipeline for RTL code and netlist.}
    \label{fig:pipeline}
    \end{figure}

    \begin{figure*}[ht]
    \centering
    \includegraphics[width=0.9\textwidth]{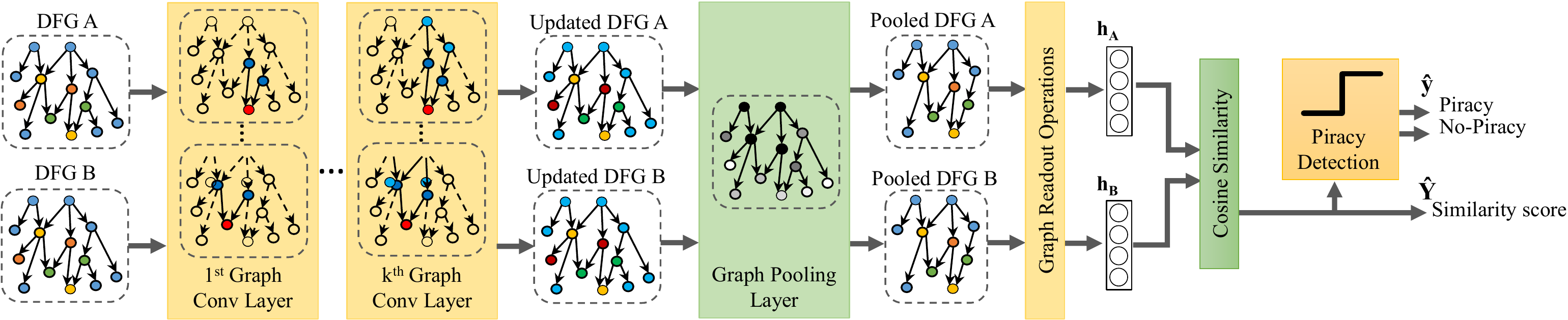}
    \vspace{-1em}
    \caption{The overall architecture of \textit{GNN4IP} for hardware IP piracy detection.}
    \vspace{-2em}
    \label{fig:model}
    \end{figure*}

\vspace{-0.5em}
\subsection{Hardware Data Flow Graph Extraction}
\label{subsec:graphextraction}

Hardware design is non-Eulicidian structural data that shares similar properties with a graph. We generate DFGs from either RTL code or gate-level netlist as the first step to model it. The DFG is a rooted directed graph illustrating the computation structure and the data flow from the circuit's output signals (the root nodes) to the input signals (the leaf nodes). It is defined as graph $G = (V, E)$ where $V =\{v_1, v_2,..., v_n\}$  is the vertices set and each node $v_i$ represents a signal, constant value, or operations such as concatenation, branch, Boolean operators, etc. We define set of directed edges $E ={e_{ij}}$ for all $i, j$ such that $e_{ij} \in E$ if the operation $v_j$ is applied on $v_i$  or the value of $v_i$ depends on the value of $v_j$.

To extract DFG, we develop an automated framework using a hardware design toolkit called Pyverilog \cite{takamaeda2015pyverilog}. Figure \ref{fig:pipeline} demonstrates our DFG generation pipeline that is consisted of five phases: preprocess, parser, data flow analysis, merge, and trim. The procedure begins with preprocessing the RTL code or gate-level netlist in Verilog format to flatten the modular codes and resolve incompatibilities and syntax errors. Afterward, the parser scans the code and produces the corresponding abstract syntax tree used by the data flow analyzer to generate a data flow tree per signal. Next, the signal's trees are merged to construct one main DFG for the whole design. Eventually, the redundant nodes and disconnected subgraphs are trimmed, and the final DFG is generated.

\vspace{-0.5em}
\subsection{Hardware IP Piracy Detection Algorithm}

Our IP piracy detection algorithm is shown in Algorithm~\ref{alg:piracyprediction}. In the algorithm, \textit{GNN4IP} refers to approximating function $f$, which can yield the inference of whether two circuits $p_1$ and $p_2$ are subject to IP piracy. Applying typical machine learning methodologies to hardware designs, which are non-euclidean in nature, usually requires feature engineering and immense expert knowledge in hardware design. Thus, we propose our scalable, automated IP piracy detection framework, \textit{hw2vec} with an architecture depicted in Figure~\ref{fig:model}.

    \begin{algorithm}[b]
    \SetAlgoLined
    \DontPrintSemicolon
    \textbf{Input:} Hardware design programs $p_1, p_2$.\;
    \textbf{Output:} A label indicating whether $p_1, p_2$ is piracy.\;
    \SetKwProg{Fn}{def}{:}{}
    \SetKwFunction{Fsgvec}{\textit{gnn4ip}}
    \SetKwFunction{Frtlvec}{\textit{hw2vec}}
    \Fn{\Frtlvec{$p$}}{
        $X, A \gets$ GraphExtraction($p$)\;
        $X^{prop}, A^{prop} \gets$ Graph\_Conv($X, A$)\;
        $X^{pool}, A^{pool} \gets$ Graph\_Pool($X^{prop}, A^{prop} $)\;
        $h_{G} \gets$ Graph\_Readout($X^{pool}$)\;
        \KwRet $h_{G}$\;
    }
    \Fn{\Fsgvec{$p_1$, $p_2$}}{
        $h_{p_1}, h_{p_2} \gets$ \textit{hw2vec}($p_1$), \textit{hw2vec}($p_2$)\;
        $\hat{Y} \gets$ {Cosine\_Sim}($h_{G_1}, h_{G_2}$)\;
        \uIf{$\hat{Y} > \delta $}{
        \KwRet $1$\;}
        \uElse{
        \KwRet $0$\;}
    }
    \textit{gnn4ip}($p_1$, $p_2$) // run the GNN4IP check.\;
    \caption{Hardware IP Piracy Detection Algorithm}
    \label{alg:piracyprediction}
    \end{algorithm}

The \textit{hw2vec} uses the DFG generation pipeline and acquires the corresponding graph $G$ for circuit $p$ in the form of $(\mathbf{X}, \mathbf{A})$ where $\mathbf{X}$ represents the initial list of node embeddings and $\mathbf{A}$ stands for the adjacency information of $G$. Next, the \textit{hw2vec} begins the message propagation phase, denoted as $\text{Graph\_Conv}$ in the algorithm, which is Graph Convolution Network (GCN)~\cite{kipf2016semi}.
In each iteration $l$ of message propagation, the node embeddings $\mathbf{X}^{l+1}$ will be updated as follows,
\begin{equation}
\vspace{-0.5em}
    \mathbf{X^{(l+1)}} = \sigma(\widehat{D}^{-\frac{1}{2}} \widehat{A} \widehat{D}^{-\frac{1}{2}} \mathbf{X^{(l)}} W^{(l)})
\end{equation}
where $W^l$ is a trainable weight used in the GCN layer. $\widehat{A} = A + I$ is the adjacency matrix of $G$ used in the layer for aggregating the feature vectors of the neighboring nodes where $I$ is an identity matrix that adds the self-loop connection to make sure the features calculated in the previous iteration will also be considered in the current iteration. $\widehat{D}$ is the diagonal degree matrix used for normalizing $\widehat{A}$. $\sigma(.)$ is the activation function such as Rectified Linear Unit (ReLU). Here, we denote the initial node embedding as $X^{(0)}$ and initialize each node embedding $\mathbf{X}^{(0)}_{i}$, $\forall i \in V$, by directly converting the node's name to its corresponding one-hot vector. We denote the final propagation node embedding $\mathbf{X}^{(l)}$ as $\mathbf{X}^{prop}$, and denote the corresponding adjacency matrix as $\mathbf{A}^{prop}$.

Once propagated the information on $G$, the resultant node embedding $\mathbf{X}^{prop}$ is further processed with an attention-based graph pooling layer $\text{Graph\_Pool}$.
Denote the collection of all node embeddings of $G$ after passing through $L$ layers of GCN as $\mathbf{X}^{prop}$. 
The $\mathbf{X}^{prop}$ is passed to a self-attention graph pooling layer that learns to filter out irrelevant nodes from the graph, creating the pooled set of node embeddings $\mathbf{X}^{pool}$ and their edges $\mathbf{A}^{pool}$. 
In this layer, we use a graph convolution layer to predict the scoring $\alpha = \mathbf{SCORE}(\mathbf{X}^{prop}, \mathbf{A}^{prop})$ and use $\alpha$ to perform \textit{top-k} filtering over the nodes in the DFG~\cite{lee2019self}.

    \begin{figure*}[t]
    \centering
    \includegraphics[width=0.98\textwidth, height=0.2\textheight]{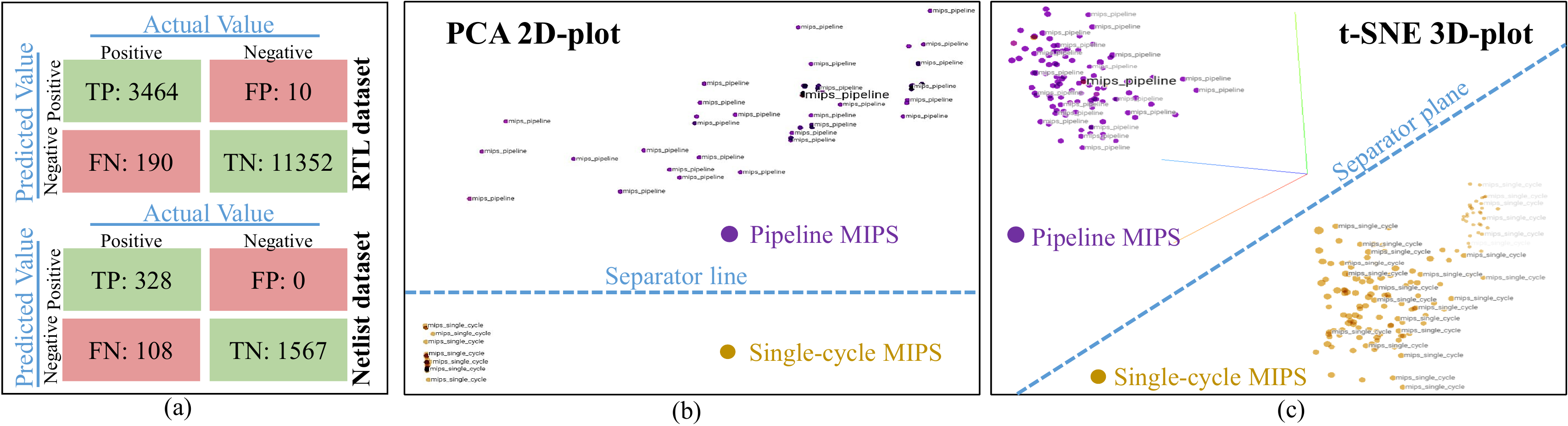}
    \vspace{-1em}
    \caption{(a) Confusion matrices for IP piracy detection, (b) \textit{hw2vec} embedding visualization using PCA, and (c) \textit{hw2vec} embedding visualization using t-SNE.}
    \vspace{-2em}
    \label{fig:embedding}
    \end{figure*}
    
Then, the $\text{Graph\_Readout}$ in our algorithm aggregates the node embeddings $\mathbf{X}^{pool}$ to acquire the graph-level embedding $\mathbf{h}_{G}$ for the DFG $G$ using this formula $\mathbf{h}_{G} = \textbf{READOUT}(\mathbf{X}^{pool})$.
The \textbf{READOUT} operation can be either summation, averaging, or selecting the maximum of each feature dimension over all the node embeddings, denoted as \textit{sum-pooling}, \textit{mean-pooling}, or \textit{max-pooling} respectively. 
Lastly, \textit{hw2vec} returns the embedding $\mathbf{h}_{G}$ of each hardware.

The \textit{gnn4ip} utilizes \textit{hw2vec} to transform $p_1$ and $p_2$ into the corresponding DFG embeddings, denoted as $h_{p_1}$ and $h_{p_2}$.
Then, it calculates the cosine similarity of $h_{p_1}$ and $h_{p_2}$ to produce the final IP piracy prediction, denoted as $\hat{Y} \in [-1, 1]$. 
The formula can be written as follows,
\vspace{-0.5em}
\begin{equation}
\vspace{-0.5em}
    \hat{Y} = \text{Cosine\_sim}(h_{p_1}, h_{p_2}) = 
    \frac{h_{p_1} \cdot h_{p_2}}{|h_{p_1}||h_{p_2}|} 
\end{equation}
Finally, our \textit{gnn4ip} utilizes predefined decision boundary $\delta$ and $\hat{Y}$ to judge whether two programs $p_1$ and $p_2$ are a piracy to one another as described in Algorithm~\ref{alg:piracyprediction} and to return the results of IP piracy detection using a binary label (0 or 1).

As both \textit{gnn4ip} and \textit{hw2vec} include several trainable parameters, we need to train these parameters for IP piracy detection via computing the cosine embedding loss function, denoted as $H$, between true label $Y$ and the predicted label $\hat{Y}$. 
The calculation of loss can be described as follows,
\vspace{-0.5em}
\begin{equation}
\vspace{-0.5em}
    H(\hat{Y}, Y) = \left \{
    \begin{array}{ll}
         1 - \hat{Y}, & \text{if $Y = 1$}\\
         max(0, \hat{Y}-\text{margin}) &  \text{if $Y = -1$}
    \end{array}
    \right.
\end{equation} 
where the margin is constant to prevent the learned embedding to be distorted (always set to 0.5 in our work). 
Once the model is trained, our algorithm uses the $\hat{Y}$ and a decision boundary $\delta$ to make the final judgement of IP piracy.

\vspace{-0.5em}
\section{Evaluation}

In \textit{hw2vec}, we use 2 GCN layers with 16 hidden units for each layer. For the \textit{graph\_pool}, we use the pooling ratio of 0.5 to perform top-k filtering. For the \textit{graph\_readout}, we use \textit{max-pooling} for aggregating node embeddings of each graph. In training, we apply dropout with a rate of 0.1 after each GCN layer. We train the model using the batch gradient descent algorithm with batch size 64 and the learning rate to be 0.001.

\vspace{-0.5em}
\subsection{Dataset}

One of the significant challenges of machine learning model development is data collection. To construct \textit{GNN4IP}, we gather RTL codes and gate-level netlists of hardware designs in Verilog format and extract their DFGs using our automated graph generation pipeline. Our collection comprises 50 distinct circuit designs and several hardware instances for each circuit design that sums up 143 netlists and 390 RTL codes. As our model works on pairs of hardware instances, we form a dataset of 19094 similar pairs and 66631 different pairs, dedicate 20\% of these 85725 pairs for testing and the rest for training.

\vspace{-0.5em}
\subsection{IP Piracy Detection Accuracy and Timing}

The \textit{GNN4IP} examines a pair of hardware designs, label it as piracy (positive) or no-piracy (negative), and outputs a similarity score in range [-1, +1] where the higher score indicates more similarity. We evaluate the model on RTL and netlist datasets, which results in the confusion matrices depicted in Figure \ref{fig:embedding}(a). We compute the IP piracy detection accuracy as the evaluation metrics, which express the correctly labeled sample ratio, true positive (TP) plus true negative (TN), to all data. The accuracy and timing results in Table \ref{tab:results} show that our model pinpoints IP piracy with high accuracy rapidly, making it scalable to large designs. The training and testing time depend on the graph size. The longer timing for netlists lies in the fact that in our dataset, the netlist DFGs with 3500 nodes on average are larger than RTL DFGs with 1000 nodes on average. We run the model on a computer with Intel Core i7-7820X CPU @3.60GHz with 16GB RAM and two NVIDIA GeForce GTX 1050 Ti and 1080 Ti GPUs and measure the timing for this computing platform.

    \begin{table}[t]
    \centering
    \caption{The \textit{GNN4IP} performance for IP piracy detection.}
    \vspace{-1em}
    \label{tab:results}
    \begin{tabular}{|C{0.8cm}|C{0.8cm}|C{0.8cm}|C{1.2cm}|C{1.3cm}|C{1.3cm}|}
    \hline
    \textbf{Dataset} &
      \textbf{\begin{tabular}[c]{@{}c@{}}Dataset\\ size\end{tabular}} &
      \textbf{\begin{tabular}[c]{@{}c@{}}\# of\\ graphs\end{tabular}} &
      \textbf{Accuracy} &
      \textbf{\begin{tabular}[c]{@{}c@{}}Train time\\ per sample\end{tabular}} &
      \textbf{\begin{tabular}[c]{@{}c@{}}Test time\\ per sample\end{tabular}} \\ \hline
    RTL &
      75855 &
      390 &
      97.21\% &
      0.577 ms &
      0.566 ms \\ \hline
    Netlist &
      9870 &
      143 &
      94.61\% &
      5.999 ms &
      5.918 ms \\ \hline
    \end{tabular}
    \end{table}

\vspace{-0.5em}
\subsection{Embedding Visualization}
  
The \textit{hw2vec} generates vectorized embedding for hardware designs and maps them to the points in the multi-dimensional space such that similar circuits are in close proximity. We visualize the \textit{hw2vec} embeddings using dimensionality-reduction algorithms such as Principal Component Analysis (PCA) and t-distributed Stochastic Neighbor Embedding (t-SNE). Figure \ref{fig:embedding} (b,c) illustrate embedding projection of 250 hardware instances for two distinct processor designs, pipeline MIPS and single-cycle MIPS, using PCA and t-SNE.

In the PCA plot, the first two principal components are depicted that express the two orthogonal directions, which maximize the variance of the projected data. t-SNE is a nonlinear machine learning algorithm that performs transformations on the data and approximate spectral clustering. We have deliberately chosen two MIPS processors with similar functionality for this experiment to harden the differentiation between them. The processors' contrast lies only in their design and specifications. According to the plots, two well-separated clusters of hardware instances are formed such that data points for the same processor design are close. It demonstrates that \textit{hw2vec} is a compelling tool to distinguish between various hardware designs. It not only considers the functionality and DFG structure but also recognizes the design.

\vspace{-0.5em}
 \subsection{Similarity Score Results}       

Our model identifies piracy based on its generated similarity score for two designs, and the decision boundary is controlled by a hyper-parameter $\delta$. We have tuned the $\delta$ to achieve maximum accuracy, but the user can adjust it to decide how much similarity is considered piracy. We calculate the similarity score in 3 cases: 1) different designs, 2) different codes with the same design, and  3) a design and its subset. For each case, 4 examples and the mean score for 50 examples are mentioned in Table~\ref{tab:similarity}. As the results present, our model successfully discriminates hardware designs since the score is very low for different designs (case1) and close to 1 for similar designs (case2). In case3, MIPS is a processor which comprises an ALU block. This relation is captured by the model and resulted in a score of approximately 0.5.

    \begin{table}[t]
    \centering
    \caption{The similarity score for a variety of hardware design pairs.}
    \vspace{-1em}
    \label{tab:similarity}
    \begin{tabular}{|C{1.1cm}|C{0.95cm}||C{1.15cm}|C{0.95cm}||C{1.1cm}|C{0.95cm}|}
    \hline
    \multicolumn{2}{|c|}{Case1}            & \multicolumn{2}{c|}{Case2}     & \multicolumn{2}{c|}{Case3}         \\ \hline
    \textbf{\begin{tabular}[c]{@{}c@{}}Circuits\\ pair\end{tabular}} &
      \textbf{\begin{tabular}[c]{@{}c@{}} Score\end{tabular}} &
      \textbf{\begin{tabular}[c]{@{}c@{}}Circuits\\ Pair\end{tabular}} &
      \textbf{\begin{tabular}[c]{@{}c@{}} Score\end{tabular}} &
      \textbf{\begin{tabular}[c]{@{}c@{}}Circuits\\ pair\end{tabular}} &
      \textbf{\begin{tabular}[c]{@{}c@{}} Score\end{tabular}} \\ \hline
    AES          & {-0.2020} & AES$_1$   & {1} & P.MIPS$_1$ & {+0.5106} \\ \cline{1-1} \cline{3-3} \cline{5-5}
    FPA          &                          & AES$_1$    &                   & ALU$_1$  &                          \\ \hline
    AES          & {-0.5240} & P.MIPS$_1$ & {+0.9939} & P.MIPS$_2$ & {+0.4965} \\ \cline{1-1} \cline{3-3} \cline{5-5}
    RS232        &                          & P.MIPS$_2$ &                   & ALU$_2$  &                          \\ \hline
    AES          & {-0.0250} & M.MIPS$_1$ & {+0.8362} & P.MIPS$_3$ & {+0.4949} \\ \cline{1-1} \cline{3-3} \cline{5-5}
    MIPS         &                          & M.MIPS$_2$ &                   & ALU$_3$  &                          \\ \hline
    FPA          & {-0.0887} & S.MIPS$_1$ & {+0.9982} & P.MIPS$_4$ & {+0.5460} \\ \cline{1-1} \cline{3-3} \cline{5-5}
    MIPS         &                          & S.MIPS$_2$ &                   & ALU$_4$  &                          \\ \hline
    Mean         &  -0.0831                        & Mean       &  +0.9571                 & Mean     & +0.5342                  \\ \hline
    \multicolumn{6}{l}{$^{\mathrm{*}}$FPA: Floating Point Adder, P.MIPS: Pipeline MIPS, M.MIPS: Multi-cycle} \\
    \multicolumn{6}{l}{MIPS, S.MIPS: Single-cycle MIPS, $X_i$: $i_{th}$ instance of hardware X.}
    \end{tabular}
    \end{table}

\vspace{-0.5em}
\subsection{Piracy Detection in Obfuscated Netlists}
\label{sec:obfuscation}

To further evaluate our model, we test it on a dataset of ISCAS'85 benchmarks, and their obfuscated instances in the gate-level netlist format, derived from TrustHub \cite{tehranipoor2016trusthub}. Obfuscation complicates the circuit and confuses reverse engineering but does not change the behavior of the circuit. Our model recognizes the similarity between the circuits despite the obfuscation because it learns the circuit's behavior. We test this capability in this experiment by comparing each benchmark with its obfuscated instances and computing each benchmark's average similarity score, presented in Table \ref{tab:obfuscated}. In the experimental results, all the similarity scores are very close to 1. It means \textit{GNN4IP} can identify the original IP in the obfuscated design 100\% of the time and is resilient against the attacks when the adversary manipulates the design to conceal the stolen IP. 
Furthermore, we assess our model on the pairs of different netlist instances, and the resultant average similarity is very low and closer to -1. It demonstrates that \textit{GNN4IP} is potent in differentiating the varied designs at the netlist level.

\vspace{-0.5em}
\subsection{Comparison with Rival Methods}   

The current state-of-the-art IP piracy detection method is watermarking. The concept of accuracy is not defined for it, and another metric called the probability of coincidence ($P_c$) is used. It declares the probability that a different designer inserts the same watermark and depends on the watermark signature size. Although the quantitative comparison with watermarking is not plausible, the false-negative rate provides similar intuition in machine learning. The state-of-the-art \cite{rai2019hardware} outperforms its previous rival algorithms by reporting $P_c$= 1.11$\times10^{-87}$ with the cost of adding 0.13\% to 26.12\% overhead to design. Our model false-negative rate is zero for netlist and  6.65$\times10^{-4}$ for RTL dataset which is very low and acceptable. Compared to \cite{rai2019hardware}, our model have the paramount advantages of zero overhead and resiliency over attacks against watermarking. Moreover, our model is powerful enough to recognize the similarity between designs despite obfuscation.

To the best of our knowledge, we are the first to model hardware as a graph for IP piracy detection. \cite{fyrbiak2019graph} utilizes graph similarity algorithm to assess obfuscation, similar to Section \ref{sec:obfuscation}. Due to different datasets exact comparison is not feasible. However, our similarity scores on obfuscation assessment notably better identify the original IP in the obfuscated one and distinguish the different designs. Their computation time is in order of minutes and significantly slower due to the graph similarity algorithm's high complexity and lack of scalability. 

    \begin{table}[t]
    \centering
    \caption{The similarity scores for obfuscated ISCAS'85 benchmarks.}
    \vspace{-1em}
    \label{tab:obfuscated}
    \begin{tabular}{|C{0.9cm}|C{4.1cm}|C{1cm}|C{1cm}|}
    \hline
    \textbf{Circuit} &
      \textbf{Circuit Function} &
      \textbf{\# of circuits} &
      \textbf{Score} \\ \hline
    c432  & 27-channel interrupt controller          & 24 &   +0.9998    \\     \hline
    c499  & 32-bit single error correcting           & 23 &   +0.9928    \\     \hline
    c880  & 8-bit ALU                                & 30 &   +0.9996    \\     \hline
    c1355 & 32-bit single error correcting           & 19 &   +0.9993    \\     \hline
    c1908 & 16-bit single/double error detecting     & 22 &   +0.9999    \\     \hline
    c6288 & 16 × 16 multiplier                       & 25 &   +0.9945    \\     \hline
    \multicolumn{3}{|l|}{Between benchmarks and their obfuscated instances} &
  \multicolumn{1}{c|}{+0.9976} \\ \hline \hline
    \multicolumn{3}{|l|}{Between different benchmarks} &
  \multicolumn{1}{c|}{-0.1606} \\ \hline
    \end{tabular}
    \end{table}

\vspace{-0.5em}      
\section{Conclusion}
In this paper, we propose a novel IP piracy detection methodology, called \textit{GNN4IP}, which does not have existing countermeasures shortcomings such as overhead and vulnerability to attacks. Our automated framework extracts the DFGs from RTL codes and gate-level netlist. Then, \textit{hw2vec}, our graph neural network generates embeddings for graphs according to the similarity between designs. Based on embeddings, we infer IP piracy between circuits with 96\% accuracy.  

\vspace{-0.5em}
\bibliographystyle{IEEEtran}
\bibliography{bibliography}

\begin{thebibliography}{10}
\providecommand{\url}[1]{#1}
\csname url@samestyle\endcsname
\providecommand{\newblock}{\relax}
\providecommand{\bibinfo}[2]{#2}
\providecommand{\BIBentrySTDinterwordspacing}{\spaceskip=0pt\relax}
\providecommand{\BIBentryALTinterwordstretchfactor}{4}
\providecommand{\BIBentryALTinterwordspacing}{\spaceskip=\fontdimen2\font plus
\BIBentryALTinterwordstretchfactor\fontdimen3\font minus
  \fontdimen4\font\relax}
\providecommand{\BIBforeignlanguage}[2]{{%
\expandafter\ifx\csname l@#1\endcsname\relax
\typeout{** WARNING: IEEEtran.bst: No hyphenation pattern has been}%
\typeout{** loaded for the language `#1'. Using the pattern for}%
\typeout{** the default language instead.}%
\else
\language=\csname l@#1\endcsname
\fi
#2}}
\providecommand{\BIBdecl}{\relax}
\BIBdecl

\bibitem{chen2020decoy}
J.~Chen~et al., ``Decoy: Deflection-driven hls-based computation partitioning
  for obfuscating intellectual property,'' in \emph{Design Automation
  Conference (DAC)}, 2020.

\bibitem{news1}
``Copyrights and patents, piracy and theft,'' \emph{The Washington Times},
  2018.

\bibitem{news2}
``Special 301 report,'' \emph{the United States Trade Representative}, 2017.

\bibitem{chang2016hardware}
C.-H. Chang~et al., ``Hardware ip watermarking and fingerprinting,'' in
  \emph{Secure System Design and Trustable Computing}, 2016.

\bibitem{gnn4tj}
R.~Yasaei~et al., ``Gnn4tj: Graph neural networks for hardware trojan detection
  at register transfer level,'' \emph{IEEE/ACM Design Automation and Test in
  Europe Conference (DATE'21)}, 2021.

\bibitem{fyrbiak2019graph}
M.~Fyrbiak~et al., ``Graph similarity and its applications to hardware
  security,'' \emph{IEEE Transactions on Computers}, 2019.

\bibitem{htm}
S.~Faezi~et al., ``Brain-inspired golden chip free hardware trojan detection,''
  \emph{IEEE Transaction on Information Forensics and Security (IEEE
  TIFS’21)}, 2021.

\bibitem{sina1}
S.~Faezi, R.~Yasaei, and M.~Al~Faruque, ``Htnet: Transfer learning for golden
  chip-free hardware trojan detection,'' \emph{IEEE/ACM Design Automation and
  Test in Europe Conference (DATE'21)}, 2021.

\bibitem{ashrafiamiri2018towards}
M.~AshrafiAmiri~et al., ``Towards side channel secure cyber-physical systems,''
  in \emph{Real-Time and Embedded Systems and Technologies}, 2018.

\bibitem{rai2019hardware}
S.~Rai~et al., ``Hardware watermarking using polymorphic inverter designs based
  on reconfigurable nanotechnologies,'' in \emph{2019 IEEE Computer Society
  Annual Symposium on VLSI (ISVLSI)}, 2019.

\bibitem{koushanfar2017active}
F.~Koushanfar, ``Active hardware metering by finite state machine
  obfuscation,'' in \emph{Hardware Protection through Obfuscation}, 2017.

\bibitem{xie2017delay}
Y.~Xie~et al., ``Delay locking: Security enhancement of logic locking against
  ic counterfeiting and overproduction,'' in \emph{Design Automation Conference
  (DAC)}, 2017.

\bibitem{camouflaging}
J.~Rajendran~et al., ``Security analysis of integrated circuit camouflaging,''
  in \emph{ACM conference on Computer \& communications security}, 2013.

\bibitem{patnaik2018raise}
S.~Patnaik~et al., ``Raise your game for split manufacturing: Restoring the
  true functionality through beol,'' in \emph{Design Automation Conference
  (DAC)}, 2018.

\bibitem{el2019sat}
M.~El~Massad~et al., ``The sat attack on ic camouflaging: Impact and potential
  countermeasures,'' \emph{IEEE Transactions on Computer-Aided Design of
  Integrated Circuits and Systems}, 2019.

\bibitem{wang2019reverse}
W.-C. Wang~et al., ``Reverse engineering for 2.5 d split manufactured ics,''
  \emph{Computer-Aided Design of Integrated Circuits and Systems}, 2019.

\bibitem{xie2016mitigating}
Y.~Xie and A.~Srivastava, ``Mitigating sat attack on logic locking,'' in
  \emph{conference on cryptographic hardware and embedded systems}, 2016.

\bibitem{li2017provably}
M.~Li~et al., ``Provably secure camouflaging strategy for ic protection,''
  \emph{Computer-Aided Design of Integrated Circuits and Systems}, 2017.

\bibitem{yasin2017removal}
M.~Yasin~et al., ``Removal attacks on logic locking and camouflaging
  techniques,'' \emph{IEEE Trans. on Emerging Topics in Computing}, 2017.

\bibitem{rajendran2013split}
J.~Rajendran~et al., ``Is split manufacturing secure?'' in \emph{Design,
  Automation \& Test in Europe Conference (DATE)}, 2013.

\bibitem{yasaei2020iot}
R.~Yasaei~et al., ``Iot-cad: context-aware adaptive anomaly detection in iot
  systems through sensor association,'' in \emph{IEEE/ACM International
  Conference On Computer Aided Design (ICCAD)}, 2020.

\bibitem{ashrafiamiri2020r2ad}
M.~Ashrafiamiri~et al., ``R2ad: Randomization and reconstructor-based
  adversarial defense on deep neural network,'' in \emph{Workshop on Machine
  Learning for CAD}, 2020.

\bibitem{wu2020comprehensive}
Z.~Wu~et al., ``A comprehensive survey on graph neural networks,'' \emph{IEEE
  Transactions on Neural Networks and Learning Systems}, 2020.

\bibitem{9054857}
W.~W. et~al., ``Detecting code clones with graph neural network and
  flow-augmented abstract syntax tree,'' in \emph{IEEE International Conference
  on Software Analysis, Evolution and Reengineering (SANER)}, 2020.

\bibitem{xu2017neural}
X.~Xu~et al., ``Neural network-based graph embedding for cross-platform binary
  code similarity detection,'' in \emph{SIGSAC Conference on Computer and
  Communications Security}, 2017.

\bibitem{takamaeda2015pyverilog}
S.~Takamaeda-Yamazaki, ``Pyverilog: A python-based hardware design processing
  toolkit for verilog hdl,'' in \emph{International Symposium on Applied
  Reconfigurable Computing}, 2015.

\bibitem{kipf2016semi}
T.~N. Kipf~et al., ``Semi-supervised classification with graph convolutional
  networks,'' \emph{arXiv preprint arXiv:1609.02907}, 2016.

\bibitem{lee2019self}
J.~Lee~et al., ``Self-attention graph pooling,'' \emph{arXiv preprint
  arXiv:1904.08082}, 2019.

\bibitem{tehranipoor2016trusthub}
``Trusthub,'' \emph{Available on-line: https://www.trust-hub.org}, 2016.

\end{thebibliography}

\end{document}